\documentstyle{elsart}

\def\Totem{\protect{\it {\sc Totem\/}}}
\def\plog#1{{\it plog#1}}
\def\pow#1{{\it \eta#1}}

\begin{document}
\begin{frontmatter}
\title{The neurochip \Totem{}: a case study in HEP.}
\author[infntn,unitn]{S. Dusini},
\author[infntn,unitn]{F. Ferrari}, 
\author[infntn,unitn]{I. Lazzizzera\thanksref{CA}}, 
\author[kent]{P. Lee},
\author[infntn,irst]{A. Sartori},
\author[infntn,unitn]{A. Sidoti},
\author[infntn,irst]{G. Tecchiolli}
\author[infntn,unitn]{A. Zorat}

\address[infntn]{INFN - Sezione di Padova, Gruppo Collegato di Trento
- Trento - Italy}
\address[unitn]{ Universit\`a di Trento - Trento - Italy}
\address[irst]{Istituto per la Ricerca Scientifica e Tecnologica -
Trento - Italy}
\address[kent]{University of Kent - Canterbury - United Kingdom}

\thanks[label]{Work supported by Istituto Nazionale di Fisica Nucleare
(INFN)}
\thanks[CA]{Corresponding author: Dipartimento di Fisica, 
Univ. Trento, I-38050 Povo (TN)\\ 
Tel:+39-461-881551, fax:+39-461-882014, e-mail: lazi@abacus.science.unitn.it}

\begin{abstract}
It is being  proved that the neurochip \Totem{} is a viable solution
for high quality and real time computational tasks in HEP, including
event classification, triggering and signal processing. The
architecture of the chip is based on a "derivative free" algorithm
called Reactive Tabu Search (RTS), highly performing even for low
precision weights. ISA, VME or PCI boards integrate the chip as a
coprocessor in a host computer.
This paper presents: 1) the state of the art and the next evolution
of the design of \Totem{}; 2) its ability in the Higgs search at LHC
as an example.\\
\noindent Reference number: 303
\end{abstract}
 
\end{frontmatter}

\vspace{-0.8cm}
\section{Introduction}
\vspace{-0.8cm}
Neural networks implemented as VLSI hardware are being considered as good 
candidates to solve problems of time-critical and high quality 
performance pattern recognition in High Energy Physics (HEP)
\cite{Den,Atlas,LinZ}.
The main benefit is speed, because of the massive parallel architecture. 
A cost is usually a very complex architectural structure, since common 
algorithms such as backpropagation, being derivative-based, require
high precision  computation\cite{BatTec-deriv}.\\
To gain significant improvement in this respect, Battiti and
Tecchiolli devised a "derivative-free" algorithm in the context of a
novel approach to the training problem, which is first transformed
into a {\it combinatorial optimization} task, then solved by means of
the heuristic method called {\it Reactive Tabu Search} 
(RTS)\cite{Bat-Tec-orsa,BatTec-ieee}.\\
RTS is based on the construction of {\it search trajectories}
in the space of the binary strings of length $L=N*B$, into
which $N$ weights, needed to configure a neural network, are suitably
coded using $B$ bits per weight. 
The search is intended to locate the best ``suboptimal'' minimum on a {\it
cost surface} by means of a sequence of elementary moves, each
consisting of a single bit-flip in the string of weights. When a
move is done, its inverse is forbidden for a {\it prohibition period}
of $T$ successive steps (the Glover's {\it Tabu Search}
method\cite{tabu}), allowing some  amount of {\it
diversification} in the search process.  RTS remarkably enhances such 
diversification by  dynamically adjusting the parameter $T$ through a 
simple mechanism that evaluates and {\it reacts to} the current local
shape of the cost surface.
This way it escapes rapidly from local minima and cyclings and
finds solutions even for low precision weights, moreover quite independently
from any starting point.\\
Sect. 2 is devoted to a description of the neurochip \Totem{}; 
Sect. 3 presents a new architectural design; Sect. 4 gives the results of a
sample application, namely the extraction of the Higgs events from background
in simulated data at LHC energies.

\vspace{-0.8cm}
\section{The \Totem{} chip}
\vspace{-0.8cm}
\Totem{} is a full-custom chip designed to operate as a co-processor in a host
system, carrying out the most compute-intensive operations for
RTS\cite{Pisa}. ISA and high performance PCI and VME boards have been developed
to set the coupling.
The chip includes an array of 32 parallel processors with associated
on-chip weight memory and control logic with broadcast and output
buses.\\
Pipelinings are included to speed up operations. 
A 32-bit static storage register on the output of the MACs 
allows data transfer from the neurons of a layer in a MLP net to occur
concurrently with a parallel input-multiply-accumulate operation on
all the processors.
The memory depth of 128 8-bit words allows neurons with up to 128
inputs to be implemented. Because of the sequential
access to the weights, the chip can realize different MLP topologies 
with a high degree of flexibility: the memory bank can either be
assigned to a single neuron or be partitioned among neurons on
different layers.
The sigmoid function is implemented off-chip by a RAM-based
look-up table. Up to four chips can be paralleled in each layer 
of a network.\\ 
With 250,000 transistors on a 70 $mm^2$ die manifactured in a 
1.2 $\mu m$ CMOS technology, \Totem{}  performs 1 GMAC/sec when clocked
at 30 MHz.\\
A doubling in the processor density and higher operational speed will
be obtained by the transition to a 0.8 $\mu m$ CMOS technology currently in
progress.

\vspace{-0.8cm}
\section{Advances in the design of \Totem{} and the \plog{} encoding}
\vspace{-0.8cm}
Considerable percentage of the silicon real estate of the \Totem{}
chip had to be devoted to the multiplication units and to the memories
where the weights are stored.  Both areas will be
reduced by means of the already mentioned technological migration. 
However in the case of the multipliers an alternative and
complementary approach can be considered: going to a {\em logarithmic 
representation} of the feature inputs and the axon weights, so that
multiplication is replaced by addition, which is 
cheaper in terms of silicon area requirements and allows both lower
power consumption and faster clock rate. Some of the
authors (P. Lee, I. Lazzizzera, A. Sartori, G. Tecchiolli and
A. Zorat) are exploring this way indeed\cite{plog}.
The first problem is to find  a reasonable approximation to the 
{\em bin-to-log} and {\em log-to-bin} conversion, since they are quite
expensive\cite{demori}. If one defines for any positive
real number $x$ the functions
$\pow(x)= n \in\mathbf{N}$ such that $x\in [2^n,2^{n+1}[$ 
and 
$\plog(x)=\pow(x)+ x/{2^{\pow(x)}} -1$, 
one gets an approximation of $log_2(x)$ that has a maximum error of
only 0.0861.  
When $x = \sum_{i=0}^{W-1} b_i 2^i$ is a binary encoded positive
number with $W=2^w$ bits, then evidently $b_i =0$ for $W-1 \geq i >
\pow(x)$ and $b_{\pow(x)}=1$. Writing 
$plog(x) = \sum_{i=-f}^{w-1} p_i 2^i$, 
it follows that the bits $p_{w-1}, p_{w-2}, \ldots, p_{0}$ are the binary 
encoding of $\pow(x)$ and the bits
$p_{-1}, p_{-2}, \ldots , p_{-f}$ are given by $b_{\pow(x)-1}, b_{\pow(x)-2}, 
\ldots b_{\pow(x)-f}$ respectively. Clearly $f$ is an integer parameter
stating a truncation (quantization error).\\
This way one gets the basis of a sign-magnitude, fixed-point 
{\em plog-encoding} of an integer $x \in [-2^W-1, 2^W-1]$ with $1+w+f$ bits: 
1 bit (given by $b_W$) for the sign; $w$ bits encoding $\pow(x)$ (the
integer part);  $f$ bits (given by 
$b_{\pow(x)-1} b_{\pow(x)-2} \ldots b_{\pow(x)-f}$) for the 
fractional part ($x=0$ is coded in a particular way).
The total error
includes the quantization ($\sim 2^{-f}$) and the approximation
to $log_2$ ($\le 0.0861$): it amounts at most to a $10\%$.
When applying the \plog{} encoding to neural nets, the multiplier
stage of a neuron is replaced by an adder and a plog-to-bin unit. In
such a \plog{} based architecture the RTS training method for a Multi Layer
Perceptron (MLP) is applied without modifications.
The above estimated error of at most $10\%$ turns out to be the
same that \Totem{} owes to a 4-bit weight setup in the conventional 
multiplier architecture: with such low precision weights, however,
adequate solutions for many problems\cite{BatTec-ieee} are still
obtained. The point is that, assuming the same fabrication technology, the
area of the multiplication blocks are reduced by
a factor 10, with a reduction in power consumption by a factor 12 and
an increase in computational speed by a factor 3.
The reduction in power consumption can be exploited to increase 
both the number of processors per die and the operating speed. These 
figures pave the way to new implementations with high performance
factors at the same fabrication costs. 
As an example the fabrication of a neural processor hosting 
hundreds of neurons running at a 
reasonable 100 MHz clock rate is feasible within a couple of years. 
With such a processor, triggering tasks requiring neural nets of 
approximately 100 neurons can easily substain input rates
of the order of $10^7$ events per second, thus making its use 
possible even in the most time critical experiments, such as LHC.

\vspace{-0.8cm}
\section{Higgs search: observables and performance of \Totem{}}
\vspace{-0.8cm}
\Totem{} has been tested in the discrimination of Higgs events from
background at LHC energies using simulation data obtained by the 
PYTHIA/JETSET Monte Carlo code.\\
 Arbitrarily we assume the Higgs mass to be 
$M_H=200\enskip GeV/c^2$, just above the threshold for the creation of two
real $Z$'s\cite{LW}.  In this case the dominant production mechanism is the
gluon--gluon fusion and the best decay channel for its identification
is the so--called {\it gold plated channel}:\\
\hspace*{.5cm} $pp\rightarrow H X\rightarrow ZZ\rightarrow 4\mu X$.\\
whose cross section is $2.84 \times 10^{-12}\ mb$ as computed by the
Pythia MC code.\\
We provide the two expected main backgrounds according to the actual
top quark mass ($M_t = 175$ GeV\cite{mtop}): \\
\hspace*{.5cm} $ p~ p \to t~ {\bar t}~ X \to \mu^+\mu^-\mu^+\mu^- X^\prime $\\
with 4 muons produced by semileptonic decays of the top and antitop;\\
\hspace*{.5cm} $ p~ p \to Z^0 ~ b ~{\bar b}~ X\to \mu^+\mu^-\mu^+\mu^- X^\prime $\\
with a muon pair produced by $Z^0$ decay and the other one by
semileptonic $b$ and ${\bar b}$ decays. These two backgrounds have
cross-sections respectively of $7.84 \times 10^{-9}\ mb$ and $6 \times 10^{-9}
\ mb$ as computed by the Pythia MC code. \\ 
We order the final muons according to the magnitude of their
transverse momenta and use the following ten variables as physical
observables :\\
\indent ($X_1-X_4$) the transverse momentum of the four muons;\\
\indent ($X_5-X_8$) the invariant masses of the four $\mu^+\mu^-$ pairs;\\  
\indent ($X_9$) the four muons invariant mass;\\
\indent ($X_{10}$) the hadron multiplicity of the hard jets,
discriminated according to the $K_{\perp}$ {\em Clustering algorithm}
for hadron-hadron collisions \cite{Catt}.\vspace*{1.mm}\\
\hspace*{.5cm} \Totem{} has been trained using a sample of 4000 Higgs events,
mixed with 2000 of each of the backgrounds. The test set, 
totally different from the training one, consisted of $N_H$ =  2000
Higgs events mixed with about 360,000 $t\bar t$ and 270,000
$Zb\bar b$ event samples (thus respecting only the ratio between the cross
sections of the two backgrounds). Some results are listed in Table
\ref{table1}, where ${N_H^c}$ is the number the events correctly
classified as Higgs, ${N_B^c}$ is the number of the events wrongly
classified as Higgs and $\delta$ is the interval amplitude within
which the classification of an Higgs is assumed certainly
correct, in units corresponding to $1/8192$ of the gap between the {\em
truth value} of an Higgs event and the {\em truth value} of
a background event. Efficiency ($N_H^c/N_H$) and purity
($N_H^c/(N_H^c+N_B^c)$) are also shown, linearly extrapolated to a
number of background events in a ratio with 2000 Higgs events as required 
by the exitimated cross sections above.\\

\vspace{-0.1cm} 
\begin{table}[hb1]
\begin{center}
\begin{tabular}{|l|r|r|r|r|}
\hline 
$\delta$ &\ \ \ \ \ \ \ ${N_H^c}$ &\ \ \ \ \ \ \ ${N_B^c}$ &\ \ \ \ \
\ \ $eff.$ &\  ${extrap.\ pur.}$ \\
\hline
\vspace{-0.2cm}
1 &753 &31 &0.37 &0.61\\
\vspace{-0.2cm}
2 &1102 &55 &0.55 &0.56\\
\vspace{-0.2cm}
5 &1228 &95 &0.61 &0.45\\
\vspace{-0.2cm}
10 &1498 &172 &0.74 &0.36\\
100 &1863 &848 & 0.93 &0.12\\
\hline 
\end{tabular}
\end{center}
\caption{}
\label{table1}
\end{table}

\end{document}